\documentclass{ws-ijmpe}
\usepackage[super,compress]{cite}
\usepackage{hyperref}

\begin{document}

\markboth{Cervia, Patwardhan, Balantekin}{On symmetries of Hamiltonians describing systems with arbitrary spins}


\title{On symmetries of Hamiltonians describing systems with arbitrary spins}

\author{Michael J. Cervia}

\address{Department of Physics, University of Wisconsin, Madison, Wisconsin 53706 USA
cervia@wisc.edu}

\author{Amol V. Patwardhan}

\address{Department of Physics, University of Wisconsin, Madison, Wisconsin 53706 USA and \\
Department of Physics, University of California, Berkeley, CA 94720 USA\\
apatwardhan2@berkeley.edu}

\author{A.B. Balantekin}
\address{Department of Physics, University of Wisconsin, Madison, Wisconsin 53706 USA \\
baha@physics.wisc.edu}

\date{\today}

\maketitle


\begin{abstract}
We consider systems where dynamical variables are the generators of the SU(2) group. A subset of these Hamiltonians is exactly solvable using the Bethe ansatz techniques. 
We show that Bethe ansatz equations are equivalent to polynomial relationships between the operator invariants, or equivalently, between eigenvalues of those invariants. 
\end{abstract}


\ccode{PACS numbers: 03.65.Fd, 13.15.+g}


\section{Introduction}

For many physical systems dynamical variables can be expressed in terms of generators of the SU(2) group. Examples in nuclear physics include isospin\cite{Heisenberg:1932dw} describing proton-neutron symmetry in nuclear and hadronic physics, quasispin\cite{kerman} describing seniority-conserving pairing interactions in nuclei, pseudo-spin\cite{adler} describing decomposition of the physical spin of single particle orbitals, and neutrino flavor/mass isospin\cite{Balantekin:2006tg,Duan:2005cp} facilitating the description of collective neutrino oscillations in core-collapse supernovae and neutron star mergers. Furthermore in condensed-matter and many body physics there are a  multitude of such problems, including those where particles carrying spins are placed in different lattices. In this article we consider such systems and write a generic Hamiltonian describing one- and two-body interactions as 
\begin{equation}
\label{generic}
H = \sum_{p=1}^N \omega_p \mathbf{B} \cdot \mathbf{J}_p + \mu \sum_{p,q=1}^N g_{pq} \> \mathbf{J}_p \cdot \mathbf{J}_q 
\end{equation}
where $\mathbf{J}_p \>, p=1,\cdots N,$ are elements of mutually commuting SU(2) algebras, which can be realized in any representation of SU(2).  
The choice of the vector $\mathbf{B}$ is dictated by the physics under consideration. 
In writing Eq. (\ref{generic}) for later convenience we separated the overall strength of the interaction ($\mu$) so that the coefficients $g_{pq}$ are dimensionless. If one uses the above Hamiltonian to describe the pairing interaction between nucleons distributed over several Shell Model orbitals, one can write down the generators appearing in Eq. (\ref{generic}) as quasispin operators. If the nucleon-nucleon interaction is determined by a single parameter, which is usually taken to be the scattering length, then all $g_{pq}$ are the same and one obtains the reduced Hamiltonian
\begin{equation}
\label{reduced}
H = - \sum_{p=1}^N \omega_p J^{z}_p + \mu \sum_{\substack{p,q=1\\ p \neq q}}^N \> \mathbf{J}_p \cdot \mathbf{J}_q .
\end{equation}
Note that in the second sum we imposed the limit $p \neq q$ since the contribution $\sum_p \mathbf{J}_p^2$ is an overall constant\footnote{In the literature sometimes the condition $p < q$ is imposed which results in an overall factor of two for the second term in Eq. (\ref{reduced}).}. 
We also chose $\mathbf{B} = (0, 0, -1)$ to be consistent with the nuclear physics applications. 
(The minus sign in front of the first term can be easily eliminated by redefining the signs of the parameters of this Hamiltonian). The same Hamiltonian appears in collective neutrino oscillations: in astrophysical sites such as core-collapse supernovae and binary neutron star mergers, the sheer number of neutrinos (which can be as high as 10$^{58}$) requires taking neutrino-neutrino interactions 
into account during neutrino transport. For two flavors, neutrino flavor isospin, which rotates one flavor into the other one, can be taken as the dynamical variable\footnote{For three flavors one needs to introduce generators of the SU(3) group.}. In the limit where one averages over the angles between the momenta of interacting neutrinos (the so-called single-angle approximation), the evolution of the neutrino gas is described by the Hamiltonian of Eq. (\ref{reduced}). To analyze the evolution in the adiabatic approximation\cite{Pehlivan:2011hp,Birol:2018qhx} one then needs to calculate the eigenvalues and eigenvectors of this Hamiltonian for a generic value of $\mu$, and then observe how these quantities change with $\mu$. 

Eigenvalues and eigenvectors of the Hamiltonian in Eq. (\ref{reduced}) were already constructed by Richardson\cite{rich} in the early 1960s. 
Subsequently this solution was cast into an algebraic framework\cite{gaudin} by Gaudin.  A recent review of the Gaudin techniques  as applied to collective neutrino oscillations is given in Ref.~\refcite{Balantekin:2018mpq} to which the reader is referred for further details on these techniques. Using this approach one can show that the operators 
\begin{equation}
\label{invariant}
h_p = - J^z_p + 2 \mu \sum_{\substack{q=1\\q\neq p}}^N \frac{\mathbf{J}_p \cdot \mathbf{J}_q}{\omega_p - \omega_q}
\end{equation}
are invariants of the Hamiltonian in Eq. (\ref{reduced}). These invariants mutually commute with each other for different values of $p$. Linear combinations of these invariants give us 
various conserved quantities. For example one obtains the third component of the total $\mathbf{J} = \sum_p \mathbf{J}_p$ as 
\begin{equation}
\label{totalj}
J^z = - \sum_p h_p = \sum_p J^z_p
\end{equation}
and the Hamiltonian of Eq. (\ref{reduced}) as 
\begin{equation}
\label{wsum}
H = \sum_p \omega_p h_p. 
\end{equation}
One can ask if linear combinations of the powers of $h_p$s will provide other conserved quantities. If all the generators $\mathbf{J}_p$  are realized in the 
$j_p = 1/2$ representations (as in many condensed-matter physics applications), where the eigenvalues of the  Casimir operators $\mathbf{J}_p^2 $ are $j_p(j_p+1)$, then using the property 
\begin{equation}
\sigma_i \sigma_j = \delta_{ij} + i \epsilon_{ijk} \sigma_k
\end{equation}
of the Pauli matrices, one can show that 
\begin{equation}
\label{fari1}
h_p^2 = \mu \sum_{q \neq p} \frac{h_q}{\omega_p - \omega_q} + \frac {3 \mu^2}{4} \sum_{q\neq p} \frac{1}{(\omega_p - \omega_q)^2} + \frac{1}{4}. 
\end{equation}
An explicit derivation of Eq. (\ref{fari1}) was given in Ref.~ \refcite{Dimo:2018tdu}. 
Hence at least in the case when all the $j_p =1/2$, all the higher powers of $h_p$ can be written as multi-linear combinations of $h_p$s. Here we derive similar expressions for higher values of $j_p$. 
 
\section{Eigenvalues and eigenvectors}

Here, following  Ref.~\refcite{Balantekin:2018mpq} we provide an overview of how eigenvalues and eigenvectors are constructed. Introducing the operators 
\begin{equation}
\mathbf{S}(\xi_\alpha)=\sum_{p=1}^N\frac{\mathbf{J}_p}{\omega_p-\xi_\alpha}
\label{gaudin1}
\end{equation}
the common eigenvectors of all the $h_p$s are 
\begin{equation}
| \xi_1,\ldots,\xi_\kappa \rangle = G(\xi_1,\ldots,\xi_\kappa)\bigg(\prod_{\alpha=1}^\kappa S^{+}_\alpha\bigg) |j,-j \rangle
\label{eigenstates}
\end{equation}
where $S_\alpha^+ \equiv \mathbf{S}^+(\xi_\alpha)$, $| j,-j \rangle$ is the lowest weight state of the total $\mathbf{J} = \sum_p \mathbf{J}_p$, and $G$ is a normalization constant,  
provided that the variables $\xi_{\alpha}$ satisfy the Bethe ansatz equations 
\begin{equation}
-\frac{1}{2\mu}+\sum_{p=1}^N\frac{j_p}{\omega_p-\xi_\alpha}+\sum_{\substack{\beta=1\\ \beta\neq\alpha}}^\kappa\frac{1}{\xi_\alpha-\xi_\beta}=0
\label{originalBA}
\end{equation}
for $\alpha=1,\ldots,\kappa$. Different $\kappa$ values obtained from solving Eq. (\ref{originalBA}) give rise to different eigenstates, orthogonal to each other. The eigenvalues of 
$h_p$, 
\begin{equation}
h_p | \xi_1,\ldots,\xi_\kappa \rangle = \epsilon_p | \xi_1,\ldots,\xi_\kappa \rangle ,
\end{equation}
are given by 
\begin{equation}
\epsilon_p=2\mu\sum_{\substack{q=1\\q\neq p}}^N\frac{j_pj_q}{\omega_p-\omega_q}+j_p-2\mu j_p \sum_{\alpha=1}^\kappa\frac{1}{\omega_p-\xi_\alpha}. 
\label{chargeValues}
\end{equation}
The form of the last term above motivates defining the quantity 
\begin{equation}
\Lambda(\lambda)=\sum_{\alpha=1}^\kappa\frac{1}{\lambda-\xi_\alpha},
\label{Lambda}
\end{equation}
using which Eq. \eqref{originalBA} can be cast into the form of a differential equation
\begin{equation}
\Lambda(\lambda)^2+\Lambda'(\lambda)-\frac{1}{\mu}\Lambda(\lambda)=\sum_{q=1}^N2j_q\frac{\Lambda(\lambda)-\Lambda(\omega_q)}{\lambda-\omega_q} ,
\label{LambdaODE}
\end{equation}
where prime denotes derivative with respect to $\lambda$. 
One approach to solving this equation is to evaluate $\Lambda (\lambda)$ and its derivatives\footnote{This idea was pursued by many authors in the literature, here we adopt the presentation given in  Refs.~\refcite{Babelon:2007td,Faribault:2011rv}.}  with respect to $\lambda$ only for $\lambda = \omega_q$. One gets  
\begin{equation}
\label{j12}
\Lambda^2 (\omega_q) + (1-2 j_q) \Lambda' (\omega_q) - \frac{1}{\mu} \Lambda (\omega_q)= 2 \sum_{p \neq q}  j_p 
\frac{\Lambda(\omega_q ) - \Lambda (\omega_p)}{\omega_q - \omega_p} .
\end{equation}
If all the SU(2) algebras are realized in the $j_q=1/2$ representation, the derivative term vanishes and one gets a set of algebraic equations with unknowns $\Lambda (\omega_p) $.  For higher $j_q$ values one needs to take higher derivatives. 
For example taking the second derivative of the Eq. (\ref{LambdaODE}) with respect to  $\lambda$ and then substituting $\lambda = \omega_q$ 
one obtains 
\begin{eqnarray}
\label{j1}
2 \Lambda (\omega_q) \Lambda'(\omega_q) &+& (1-j_q)\Lambda''(\omega_q) - \frac{\Lambda'(\omega_q)}{\mu}  \nonumber \\ 
&=& 2 
\sum_{p \neq q} j_p \left[ \frac{\Lambda'(\omega_q)}{\omega_q - \omega_p} - \frac{\Lambda(\omega_q) - \Lambda(\omega_p)}{(\omega_q - \omega_p)^2} \right]  
\end{eqnarray}
The second term vanishes for $j_q =1$ and again one gets a set of algebraic equations with unknowns $\Lambda (\omega_q) $ and 
$\Lambda' (\omega_p) $. This procedure can be repeated to obtain sets of algebraic equations for all values of $j_q$. 

\section{Higher powers of conserved quantities}

Since Eq. (\ref{chargeValues}) can be rewritten as 
\begin{equation}
\epsilon_p=2\mu\sum_{\substack{q=1\\q\neq p}}^N\frac{j_pj_q}{\omega_p-\omega_q}+j_p-2\mu j_p \Lambda (\omega_p)
\label{chargeValues2}
\end{equation}
and $h_p$ can be expanded as 
\begin{equation}
\label{hpexpansion}
h_p = \sum_{\kappa} \sum_{d_{\kappa}} \epsilon_p (\xi_1, \xi_2, \cdots, \xi_{\kappa})
 | \xi_1, \xi_2, \cdots, \xi_{\kappa} \rangle \langle  \xi_1, \xi_2, \cdots, \xi_{\kappa} | ,
\end{equation}
where $d_{\kappa}$ is a degeneracy parameter which distinguishes different solutions of the Bethe ansatz with the same value of $\kappa$, identities such as Eq. (\ref{j12}) and (\ref{j1}) 
also hold for $h_p$ with suitable modifications. 

To obtain explicit expressions it is convenient to locally continue $\epsilon_p$ to a function of $\lambda$ close to $\omega_p$ as
\begin{equation}
\tilde{\epsilon}_p (\lambda)=2\mu\sum_{\substack{q=1\\q\neq p}}^N\frac{j_pj_q}{\lambda-\omega_q}+j_p- 2 \mu j_p\Lambda(\lambda),
\label{epsilon}
\end{equation}
and define $n^\mathrm{th}$ derivative parameters $\epsilon_p^{(n)}$ as
\begin{equation}
\epsilon_p^{(n)} \equiv \lim_{\lambda\to\omega_p}\tilde{\epsilon}_p^{(n)}(\lambda)=(-1)^nn!2\mu\sum_{\substack{q=1\\q\neq p}}^N\frac{j_pj_q}{(\omega_p-\omega_q)^{n+1}}-2\mu j_p\Lambda_p^{(n)},
\label{epsilonDeriv}
\end{equation}
where $\Lambda_p = \Lambda (\omega_p)$, and $\Lambda_p^{(n)}$ is the $n$-th derivative of $\Lambda(\lambda)$ as ${\lambda\to\omega_p}$. 
Next we rewrite Eq. \eqref{LambdaODE} as $N$ equations in $\tilde{\epsilon}_p (\lambda)$ with $\lambda\sim\omega_p$ to obtain 
\begin{equation}
\tilde{\epsilon}^2_p(\lambda)+2\mu j_p\bigg[2j_p\frac{\tilde{\epsilon}_p (\lambda)-\epsilon_p}{\lambda-\omega_p}- \tilde{\epsilon}_p' (\lambda)\bigg]=\sum_{\substack{q=1\\q\neq p}}^N\tilde{C}_{pq}(\lambda)\epsilon_q+ \tilde{K}_p(\lambda),
\label{EpsilonODE1}
\end{equation}
where
\begin{equation}
\tilde{C}_{pq}(\lambda) \equiv\frac{4\mu j_p^2}{\lambda-\omega_q}, \quad \tilde{K}_p(\lambda) \equiv j_p^2+4\mu^2j_p^2\sum_{\substack{q=1\\q\neq p}}^N\frac{j_q(j_q+1)}{(\lambda-\omega_q)^2} .
\label{EpsilonODE2}
\end{equation}
Differentiating Eq. \eqref{EpsilonODE1} $n$ times and taking the limit $\lambda\to\omega_p$, we obtain the algebraic equations
\begin{equation}
\sum_{k=0}^n\binom{n}{k}\epsilon_p^{(n-k)}\epsilon_p^{(k)}+2\mu j_p\frac{2j_p-(n+1)}{(n+1)!}\epsilon_p^{(n+1)}=\sum_{\substack{q=1\\q\neq p}}^NC_{pq}^{(n)}\epsilon_q+K_p^{(n)}, 
\label{chargeBAn1}
\end{equation}
where 
\begin{equation}
C_{pq}^{(n)} \equiv  \tilde{C}_{pq}^{(n)}(\omega_p) = (-1)^nn! \frac{4\mu j_p^2}{(\omega_p-\omega_q)^{n+1}}, 
\label{chargeBAn2a}
\end{equation}
and
\begin{equation}
K_p^{(n)} \equiv \tilde{K}_{p}^{(n)}(\omega_p)=j_p^2\delta_{n0}+ (-1)^n(n+1)!\sum_{\substack{q=1\\q\neq p}}^N\frac{4\mu^2j_p^2j_q(j_q+1)}{(\omega_p-\omega_q)^{n+2}} .
\label{chargeBAn2}
\end{equation}

Note that, for a given $j_p > 1/2$, we may obtain recurrence relations from Eq. \eqref{chargeBAn1} with $n<2j_p-1$, relating derivative parameters $\epsilon_p^{(n+1)}$ to lower order derivative parameters, ultimately arriving at polynomials of degree $n+1$ in $\epsilon_p$ and of strictly lower degree in $\epsilon_q$ ($q\neq p$). 
(Also note that we do not need recurrence relations for $j_p =1/2$ since the derivative term in Eq. (\ref{j12}) vanishes). 
Plugging these resulting polynomials into Eq. \eqref{chargeBAn1} with $n=2j_p-1$, we obtain a single polynomial equation of degree $2j_p+1$ in $\epsilon_p$ and degree $2j_p-1$ in $\epsilon_q$ ($q\neq p$).  Since these $\epsilon_p$ are the eigenvalues of the conserved charges, $h_p$, we expect\footnote{Since the solutions to the set of polynomial equations are $d = \prod_{p=1}^N (2j_p+ 1)$ in number, we can label each solution (including degeneracy) as $\{\epsilon_{p,i}: p = 1,\ldots,N \}$, where the index $i = 1,\ldots,d$. Indeed, since there exists a common basis of orthogonal eigenvectors $| i \rangle$ (given by Eq.~(\ref{eigenstates})), such that each of the invariants may be expressed as $h_p = \sum_i \epsilon_{p,i} | i \rangle \langle i |$, one may straightforwardly conclude that $ P(\{ h_p: p = 1,\ldots,N \}) = \sum_i P(\{ \epsilon_{p,i}: p = 1,\ldots,N \})\,| i \rangle \langle i | $, where $P$ is any multivariate polynomial in $N$ variables. The existence of $d$ solutions can be verified by solving the polynomial equations for $\mu = 0$. For $\mu > 0$, it can be shown that each solution is continuously connected to a unique $\mu = 0$ counterpart, and the number of solutions is thus preserved \cite{Garcia1979}.} 
that the invariants obey the same polynomial equations with $h_p, h_q$ replacing $\epsilon_p, \epsilon_q$. Therefore, we may also conclude that any conserved charges that may be written as polynomials of the charges $h_p$ may be reduced to polynomials of degree at most $2j_p$ in each $h_p$.

For example, taking Eq. \eqref{chargeBAn1} with $j_p=1/2$, we obtain the result given in Eq. (\ref{fari1}). In the case $j_p=1$ for a given $p$, we may combine these equations to obtain an algebraic equation for $\epsilon_p$ in terms of the other $\epsilon_q$s:
\begin{equation}
\epsilon_p^3-\bigg(\sum_{\substack{q=1\\q\neq p}}^NC_{pq}^{(0)}\epsilon_q+K_p^{(0)}\bigg)\epsilon_p+\mu\bigg(\sum_{\substack{q=1\\q\neq p}}^NC_{pq}^{(1)}\epsilon_q+ K_p^{(1)}\bigg)=0.
\label{chargeBA1}
\end{equation}
Based on the reasoning given above, we similarly have the operator equations
\begin{equation}
h_p^3-\bigg(\sum_{\substack{q=1\\q\neq p}}^NC_{pq}^{(0)}h_q+K_p^{(0)}\bigg)h_p+\mu\bigg(\sum_{\substack{q=1\\q\neq p}}^NC_{pq}^{(1)}h_q+ K_p^{(1)}\bigg)=0.
\label{chargeBAop1}
\end{equation}
Note that other $j_q$ can still take arbitrary values (in multiples of $1/2$). This does not affect the form of Eq.~\eqref{chargeBA1}, with the only change being in the terms of $K_p^{(n)}$. 
However, corresponding equations for other $\epsilon_q (q \neq p)$ may not have the same order as the equation for $\epsilon_p$. 

For more examples, we obtain for $j_p=3/2$
\begin{flalign}
\epsilon_p^4&-\frac{10}{9}\bigg(\sum_{\substack{q=1\\q\neq p}}^NC_{pq}^{(0)}\epsilon_q+K_p^{(0)}\bigg)\epsilon_p^2+\frac{8}{3}\mu\bigg(\sum_{\substack{q=1\\q\neq p}}^NC_{pq}^{(1)}\epsilon_q+ K_p^{(1)}\bigg)\epsilon_p 
\nonumber \\
&+\frac{1}{9}\bigg(\sum_{\substack{q=1\\q\neq p}}C_{pq}^{(0)}\epsilon_q+K_p^{(0)}\bigg)^2-2\mu^2\bigg(\sum_{\substack{q=1\\r\neq p}}C^{(2)}_{pq}\epsilon_q+K_p^{(2)}\bigg)=0,
\label{chargeBA32}
\end{flalign}
and for $j_p=2$ we get 
\begin{eqnarray}
\epsilon_p^5 &-& \frac{6}{5} \bigg(\sum_{\substack{q=1\\q\neq p}}^N C^{(0)}_{pq} \epsilon_q+K_p^{(0)}\bigg)\epsilon_p^3
   +\frac{26}{5} \mu \bigg(\sum _{\substack{q=1\\q\neq p}}^N C^{(1)}_{pq} \epsilon_q+K_p^{(1)}\bigg)\epsilon_p^2 \nonumber \\
   &+& \frac{1}{5}\bigg[\bigg(\sum_{\substack{q=1\\q\neq p}}^N C^{(0)}_{pq} \epsilon_q+K_p^{(0)}\bigg)^2 
   -48 \mu ^2 \bigg(\sum_{\substack{q=1\\q\neq p}}^N C^{(2)}_{pq} \epsilon_q+K_p^{(2)}\bigg)\bigg]\epsilon_p \nonumber \\
   &-&\frac{2}{5} \mu\bigg(\sum_{\substack{q=1\\q\neq p}}^N C^{(0)}_{pq} \epsilon_q+K_p^{(0)}\bigg) 
   \bigg(\sum_{\substack{q=1\\q\neq p}}^N C^{(1)}_{pq} \epsilon_q+K_p^{(1)}\bigg) \nonumber \\
   &+& \frac{16}{5} \mu ^3 \bigg(\sum_{\substack{q=1\\q\neq p}}^N C^{(3)}_{pq} \epsilon_q+K_p^{(3)}\bigg) =0.
\label{chargeBA2}
\end{eqnarray}

It is straightforward to verify the solutions to these equations in the limit where the neutrino self-interaction vanishes, i.e., $\mu \to 0$. From the definition of the operators $h_p$ given in Eq.~\eqref{invariant}, it can be seen that the eigenvalues in the $\mu \to 0$ limit are $\epsilon_p = -j_p,\ldots,j_p$ (in increments of $1$). For example, in this limit, the equation for $j_p = 1$, Eq.~\eqref{chargeBA1}, reduces to $\epsilon_p^3 - \epsilon_p = 0$, yielding the solutions $\epsilon_p = -1,0,1$. The equation for $j_p = 3/2$, Eq.~\eqref{chargeBA32}, reduces to $\epsilon^4 - \frac 52 \epsilon_p^2 + \frac 9{16} = 0$, yielding $\epsilon_p = -3/2,-1/2,1/2,3/2$, as expected (note that, as $\mu \to 0$, $C_{pq}^{(n)} = 0$ and $K_p^{(n)} = j_p^2 \delta_{n0}$). In fact, it is possible to show this in the general case as well, starting from the original form of the Bethe Ansatz equations.

In the $\mu \to 0$ limit, the roots of Eq.~\eqref{originalBA} are known to converge to the $\omega_p$s. A root $\xi_\alpha$ which converges to a particular $\omega_p$ could then be expressed as a power series in $\mu$, i.e., $\xi_\alpha = \omega_p + \mu\,x_{\alpha,p} + \mathcal O(\mu^2)$, giving the following Bethe Ansatz equations at lowest order in $\mu$:
\begin{equation}
\label{BAmu0}
-\frac 12 - \frac{j_p}{x_{\alpha,p}} + \sum_{\substack {\beta = 1 \\ \beta \neq \alpha} }^{\kappa_p} \frac{1}{x_{\alpha,p} - x_{\beta,p}} = 0,
\end{equation}
where $\kappa_p$ is the number of roots converging to a particular $\omega_p$. It is possible~\cite{Araby:2012ru} to show that $x_{\alpha,p}$ are the roots of generalized Laguerre polynomials $L_{\kappa_p}^{-1-2j_p}(x)$. Then one can use the definition from Eq.~\eqref{chargeValues} to infer that
\begin{equation}
\lim_{\mu \to 0} \epsilon_p = j_p + 2j_p \sum_{\alpha = 1}^{\kappa_p} \frac{1}{x_{\alpha,p}} = j_p - \kappa_p
\end{equation}
where the last equality can be obtained by summing Eq.~\eqref{BAmu0} over the $\kappa_p$ roots. Knowing that the number of roots that converge to a particular $\omega_p$ can range from $0$ to $2j_p$ (corresponding to the number of times a raising operator can act on the state $|j_p,-j_p \rangle$), this reproduces the expected eigenvalues $\epsilon_p = -j_p,\ldots,j_p$.

\section{Conclusions}

We showed that invariants of the Hamiltonian in Eq. (\ref{reduced}), the quantities designated in this paper as $h_p$, satisfy certain polynomial relations amongst themselves. The same relations also hold for the eigenvalues. Since these relationships are obtained using the Bethe ansatz equations, solving them is completely equivalent to solving Bethe ansatz equations. This fact was also emphasized in Ref.~\refcite{Faribault:2018tao} in the context of $j_p=1/2$. 
Here we showed it to be true for all values of $j_p$.

\section*{Acknowledgements}

This paper is dedicated to the memory of Ernest Henley. This work was supported in part by the US National Science Foundation Grants No.  PHY-1630782 and PHY-1806368.

\end{document}